\documentclass[doublecol]{my-epl2}

\usepackage{graphicx}

\usepackage{ulem}  

\usepackage{graphicx}
\usepackage{amsmath}
\usepackage{amssymb}
\usepackage{yfonts}
\usepackage{slashed} 

\usepackage{hyperref}

\usepackage{marvosym}
\usepackage{wasysym}

\newcommand{\myskip}[1]{}

\newcommand{\cO}{{\cal O}}

\newcommand{\gam}{\gamma}
\newcommand{\mn}{{\mu\nu}}

\newcommand{\ext}{{\rm ex}}
\newcommand{\inn}{{\rm in}}

\newcommand{\cc}{{\rm cc}}

\newcommand{\vB}{{\bf B}}
\newcommand{\vE}{{\bf E}}

\newcommand{\diag}{{\rm diag}}
\newcommand{\cosm}{{0}}
\newcommand{\qad}{\hspace{1mm}}

\newcommand{\cCc}{{\cal C}}
\newcommand{\veps}{\varepsilon}

\newcommand{\mmin}{{\hspace{0.2mm}-\hspace{0.2mm}}}
\newcommand{\pplus}{{\hspace{0.2mm}+\hspace{0.2mm}}}

\newcommand{\iss}{{\hspace{0.2mm}=\hspace{0.2mm}}}

\newcommand{\onedot}{\,\,\,}

\newcommand{\ed}{\onedot}
\newcommand{\ednu}{{\onedot\nu}}

\graphicspath{{./}{figures/}}

\renewcommand{\em}{{{\rm em}\,}}

\newcommand{\rot}{{\rm rot}}

\renewcommand{\cc}{{\rm cc}}
\newcommand{\cs}{{\rm cc}}

\renewcommand{\d}{{\rm d}}
\newcommand{\nn}{\nonumber}

\newcommand{\Mpc}{{\rm Mpc}}

\newcommand{\BEQ}{\begin{eqnarray}}   
\newcommand{\EEQ}{\end{eqnarray}}   
\newcommand{\BEA}{\begin{eqnarray}}   
\newcommand{\EEA}{\end{eqnarray}}

\begin{document}

\title{Indefinitely flat rotation curves 
 from Mpc sized charged cocoons}

\shorttitle{Indefinitely flat rotation curves  from Mpc sized charged cocoons}

\author{Theodorus Maria Nieuwenhuizen}

\shortauthor{T. M. Nieuwenhuizen}

\institute{Institute for Theoretical Physics,  Science Park 904, 1098 XH Amsterdam, The Netherlands}

 \pacs{98.62.Dm}{Kinematics, dynamics, and rotation} \pacs{95.35.+d}{Dark matter} \pacs{04.20.Jb} {Exact solutions}


\newcommand{\red} {\color{red}}
\newcommand{\vac}{{{\rm v}}}

\newcommand{\rhovv}{{\rho_\cc^v}}
\newcommand{\pvv}{{p_\cc^v}}

\abstract{
Weak lensing  exhibits that rotation curves  of isolated galaxies remain flat up to Mpc scale (Mistele et al, 2024). 
 Recently we proposed that dark matter is a combination of electrostatic and vacuum energy in standard physics.
In this theory,   isolated galaxies may be  embedded in ``charged cocoons'', spheres  with $\pm 1/r^2$  charge density. 
The related circular rotation curves decay at the Mpc scale.
The analytic solution is extended to the early Universe. 
A peak in the cosmic microwave background spectrum is expected at the arcsec scale. 
The induced nanometer size of the cocoons at the Big Bang makes the initial state inhomogeneous.
}

\maketitle

\section{Introduction}

Dark matter (DM),  established by Zwicky in 1933 \cite{zwicky1933rotverschiebung}, causes the flattening of  rotation curves 
for stars and hydrogen clouds. This was evidenced  by Rubin and Ford \cite{rubin1970rotation} up to some 25 kpc for Andromeda (M31) 
and out to 122 (0.5$/h$) = 87 (0.7/$h)$ kpc in  \cite{rubin1980rotational}.
Astonishingly, a  2024 analysis of weak lensing by isolated, early and late type galaxies exhibits  ``indefinitely flat circular velocities'',
ranging out to several hundreds kpc, without  clear decline out to 1 Mpc \cite{mistele2024indefinitely}, the size of a fat galaxy cluster.

This scale of dark matter clouds is difficult to understand within the standard cosmological model $\Lambda$CDM \cite{mistele2024indefinitely},
with its NFW density profile decaying as $1/r^3$ \cite{navarro1997universal}.
It does support MOND \cite{milgrom1983modification,banik2022galactic} and MOND-like theories, see, e. g.,  
\cite{verlinde2011origin,verlinde2017emergent}, where, effectively, the Newton force decays as $1/r$.
 However, analysis of wide binaries rules out the simplest of those models at $16\sigma$, so that
MOND must be substantially modified on small scales \cite{banik2024strong} (but see  \cite{hernandez2024critical} for criticism).
This adds  to tensions of MOND in clusters  \cite{sanders2003clusters,nieuwenhuizen2017zwicky},
 the solar system \cite{desmond2024tension}  and otherwise \cite{banik2022galactic}, which  motivates to investigate different starting points.

Recently, our analysis of regular black hole interiors 
\cite{nieuwenhuizen2023exact}, the nature of dark matter \cite{nieuwenhuizen2024solution} 
and the interior of the Lorentz electron  \cite{nieuwenhuizen2024aether} have led  to assume 
that, within the standard model of particle physics and classical Einstein-Maxwell theory, 
the vacuum has a richer structure  than commonly considered. It is a state without matter the 
energy of which can flow and condense with the help of electric fields, 
so as to establish electro-vacuum energy (EVE) (or ``electro-aether energy'', EAE \cite{nieuwenhuizen2024solution}), as  the dark matter.
This is supported by  the no-show of a dark matter particle in  decades of intensive search;
by early structure formation triggered by thunders after lightnings due to charge mismatches, and various  further aspects \cite{nieuwenhuizen2024solution}.

\section{The charged cocoon}
Here we investigate whether in EVE/EAE theory,  the indefinitely flat circular rotation curves can be connected to  charged spheres, 
to be called ``charged cocoons'' (ccs).
Consider a spherically symmetric configuration of radius $R$ with charge density  $\rho_q(r)$, included charge $Q(r)=4\pi\int_0^r\d u\,u^2\rho_q(u)$
and  total charge $Q(R)=q$. In units $\hbar=c=k=1$ and  $\mu_0=1/\veps_0=4\pi$,
the electric field reads $E(r)=Q(r)/r^2$ and  the electrostatic energy density $\rho_E=E^2/8\pi$.
Neglecting for simplicity the contribution of normal matter, the cc  involves electrostatic energy $M_E$,  vacuum energy 
$M_\vac=M_E/3$  and total mass $M=4M_E/3$ for any charge configuration \cite{nieuwenhuizen2024solution}.

In the interior  (``in'', $r\le R$), we consider 
\BEQ \label{rhoqQ}
\rho_q^\inn =\frac{q } {4\pi R r^2}, \quad Q(r)=q\frac{r}{R}, \quad E(r)=\frac{Q(r)}{r^2}=\frac{q}{Rr} .
\EEQ
This leads to inverse power laws of the energy densities
\BEQ
 \label{rhoELin0}
\rho_E^\inn \equiv \frac{E^2}{8\pi}=\frac{q^2}{8\pi R^2}\frac{1}{r^2} , \qquad
 \rho_\vac^\inn  =
 \frac{q^2}{8\pi R^2}\left(\frac{1}{r^2}-\frac{1}{R^2}\right) .
\EEQ
with $\rho_\vac=\rho_E^>-\rho_E$  involving  $\rho_E^>=4\int_r^\infty \d u\,\rho_E(u)/u$ 
 \cite{nieuwenhuizen2024solution}.
In the exterior (``ex'', $r>R$), $Q(r) \iss q$, $E \iss {q}/{r^2},$ one has the standard absence of charge and vacuum energy, 
\BEQ
 \label{rhoELex0}
 \rho_q^\ext=\rho_\vac^\ext = 0 , \qquad  \rho_E^\ext =\frac{q^2}{8\pi r^4} .
\EEQ
The combination $\rho_E+\rho_\vac=\rho_E^>$ represents the EVE/EAE dark matter density.
The  included dark mass is 
\BEQ
M(r)= \left \{ \begin{matrix} \hspace{0mm} (q^2r/R^2) (1-r^2 / 6R^2),\quad \qquad  (r<R), \\ 
(4q^2/3R)(1- 3R /8r ) ,  \quad \qquad \qad \qad
(r>R)  .\end{matrix}   \right. 
\EEQ
 The  full masses of the components  follow as 
\BEQ \label{MeMvMeae}
M_E=\frac{q^2}{R},\quad M_\vac=\frac{q^2}{3R}, \qquad M=\frac{4q^2}{3R} .
\EEQ
The factor 4/3 is known from the ``problem 4/3'', the ratio of  kinematic and electrostatic mass of the Lorentz electron,
solved by Poincar\'e  in 1905 for charged spherical shells \cite{poincare1905dynamique,damour2017poincare}.
His incorporation of the vacuum is the essential piece of  our more general electro-vacuum  theory
 \cite{nieuwenhuizen2023exact,nieuwenhuizen2024solution,nieuwenhuizen2024aether}.

Again disregarding the normal matter, the circular rotation speed $v_\rot(r)=\sqrt{GM(r)/r}$ attains the profile
\BEQ \label{vrotuniflat}
v_\rot =\left \{ \begin{matrix}  v_0\sqrt{1-{r^2}/{6R^2}}, \qquad \qquad \, (r<R), \\   
v_0\sqrt{ {4R}/{3r}- {R^2}/{2r^2}} , \qquad (r> R) , \end{matrix} \right  .
\EEQ
which, holding away from the center, involves
\BEQ \label{v0=,q=}
v_0=q\frac{\sqrt{G}}{R} ,\qquad q=v_0\frac{R}{\sqrt{G} }   .
\EEQ
The profile  (\ref{vrotuniflat})  is depicted in figure 1.

\begin{figure}
\begin{center}
\includegraphics[width=8cm]{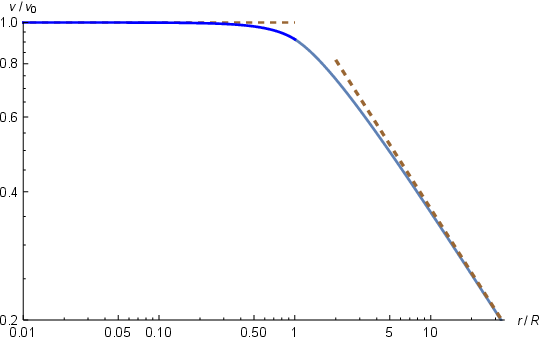}
 \end{center}
\caption{The  ``indefinitely flat'' profile  (\ref{vrotuniflat})  of rotation curves decays  beyond $R$ 
in line with  Newton's law $v\sim 1/\sqrt{r}$.  }
\label{figeen}
\end{figure}

\subsection{Physical scales}
Consider, for some $\alpha\ll 1$, a number density $n_\cs\equiv 3\alpha /4\pi  R^3$ of  charged cocoons  of typical radius $R$.
This  involves their average mass density
\BEQ
\bar \rho_\cs =n_\cs M=\frac{4\alpha q^2}{3 R^4} .
\EEQ
Assume that this constitutes a fraction $\alpha \beta^2 $ of the critical mass density $\rho_c=3H_0^2/8\pi G$,
with $\beta\approx 2$.  
This fixes $R$ to the rotation speeds,
\BEQ \label{Rval=}
R=\sqrt{\frac{8}{3}} \frac{v_0}{\beta H_0}
= 2.3 \, v_{200} \,\Mpc,
\EEQ
where $v_{200}=v_0/(200$ km/s) and  we take $H_0= 70$ km/s Mpc.
Notice that $R$  lies in the Mpc range.
The charge is
\BEQ \label{qval=}
q=\pm\sqrt{\frac{8}{3G}}\frac{v_0^2}{\beta H_0} = \pm \,3.0\,v_{200}^2 \,10^{54} \, {\rm C} ,
\EEQ 
The total EVE/EAE mass corresponds to that of a  small galaxy cluster,
\BEQ
M=\frac{4v_0^2R}{3G} 
=\frac{16}{3\beta\sqrt{6}}\frac{v_0^3}{GH_0}=
2.9  \,  v_{200}^3 \,10^{13}M_\odot . 
\EEQ

Assume next that $(5q^2/6R)/(4\pi R^3/3)$,  the dark mass density of the cc within $R$,
exceeds $\rho_c$ by a factor $\mu >1$, while normal matter has  the familiar 5\% $\rho_c$.
With 25\% of mass in $^4$He, the total number of electrons inside
$R$  is 
\BEQ
\hspace{-2mm}
N_e=\frac{7}{8}\frac{1}{ 20\mu} \frac{5q^2}{6Rm_N}
=\frac{7v_0^3}{48\sqrt{6}\beta \mu GH_0m_N}
= 3.8 v_{200}^3  10^{68} .
\EEQ
This is a fraction $5\beta^2/8\mu$ of their cosmic average number, hence $\beta=\sqrt{8\mu/5}$.
While all parameters fluctuate from cocoon to cocoon, 
we take for simplicity $\mu\sim 2.5$, $\beta\sim 2$.

The  charge $q$ corresponds to the number of uncompensated electrons within radius $r$,
\BEQ
\delta N_e(r)=-\frac{q}{e}\frac{r}{R}=- \sqrt{137} \frac{qr}{R}= \mp 8.6  v_{250}^2 \,10^{55}  \frac{r}{R},
\EEQ
so that the charge mismatch ratio results as
\BEQ \label{deltae=}
\delta_e = \frac{\delta N_e(R)}{N_e}= \mp \frac{0.92}{v_{200}} \, 10^{-13}.
\EEQ
While   $|\delta_q|\sim 10^{-13}$ is estimated for the solar neighbourhood,  
values $|\delta _q|\sim 6\, \, 10^{-15}$ have been reported for $v_{200}=5$ in 
the fat galaxy clusters A1689 and A1835   \cite{nieuwenhuizen2024solution}.

The typical electric field strength is basically universal,
\BEQ
E(r)=\frac{q}{Rr}= \frac{R}{r} \sqrt{\frac{3\mu}{5G}} H_0 \approx 10 \frac{R}{r} \,  \frac{{\rm V}}{{\rm m}} .
\EEQ
The Maxwell equation $\nabla\times \vB=\dot \vE$ sets a typical value
\BEQ
B(R)\sim H_0RE(R)= \frac{H_0v_\rot}{\sqrt{G}}  \approx  0.18 v_{200}\,  \mu{\rm G},
\EEQ
which connects to the $\mu$G scale of cosmic magnetic fields.

\section{A charged  cocoon in the expanding Universe}
As background we consider   the Friedman-Lemaire-Robertson-Walker  (FLRW) metric with scale factor  $a(t)$ and metric
$\gam_\mn=\diag(1,-a^2,-a^2r^2, -a^2r^2\sin^2\theta )$ in spherical coordinates.
With energy density $\rho_0$ and  isotropic pressure $p_0$, its stress energy tensor takes the form
 \BEQ
 T_0^\mn= (\rho_0+p_0)\delta^\mu_0\delta^\nu_0 -p_0\gam^\mn.
\EEQ
The Friedman equations connect them to the scale factor,
\BEQ\label{rho0p0}
\rho_\cosm(t)=\frac{3\dot a^2}{8\pi Ga^2} = \frac{3H^2}{8\pi G}, \qquad p_\cosm(t)=-\frac{2a\ddot a+\dot a^2}{8\pi Ga^2} .
\EEQ
We shall employ these relations  to eliminate $\rho_0$ and $p_0$,
so that a cosmological model is not needed yet.

For the temporal development of  the  cc, we consider the  generalized Schwarzschild-FLRW metric
\BEQ \label{ds2h0123}
\d s^2&=&g_\mn\d r^\mu\d r^\nu \\
&=& g_0 g_1\d t^2 -a^2\frac{g_0}{g_1} \d r^2  -a^2r^2g_2 ( \d\theta^2 \pplus \sin^2 \! \theta \,\d\phi^2), \nn 
\EEQ
with  $g_i=g_i(t,r)$ and volume element $\d V=\sqrt{-g}\d r\d\theta\d\phi$, where  $\sqrt{-g} =a^3 g_0g_2 r^2 \sin\theta$.

  The electromagnetic  potential and electric field
\BEQ
&& A_\mu= \delta^0_\mu A_0,\qquad  E=-A_0' =\frac{Q(r)}{ar^2}  , 
\EEQ
with $Q(r)$ as in (\ref {rhoqQ}),
yield the Maxwell stress energy tensor
\BEQ \label{SETMax}
T^{\mu}_{{\rm em}\,\nu} 
= \rho_E  \cCc^\mu_{1\,\nu} ,\quad 
\cCc^\mu_{1\,\nu}={\rm diag}(1,1,-1,-1), 
\EEQ
 with $\rho_E= {Q^2(r)}/ {8\pi a^4 g_0^2 r^4}$.
The case $g_2(t,r)=g_0(t,r)$  fixes the charge current  as  
$J_q^\nu=F^\mn_{\ed ;\mu}/4\pi =\delta^\nu_0  \rho_q(r) /a^3g_0^2$ and
implies the conserved global charge $\int \d V\,  J_q^0(t,r)=q$. 
The electrostatic mass emerges as  $M_E(t)=q^2/a(t)R$.

On top of its electrostatic mass, the cocoon can have surplus normal matter of baryons and radiation. 
While this is often subleading, we include it for generality.
Its stress energy tensor is $T^\mn_\cc(t,r)=T^\mu_{\cc\,\dot\nu}(t,r)g^{\dot\nu\nu}(t,r)$ with
\BEQ
T^\mu_{\cc\,\nu}=\diag(\rho_\cc ,-p_\cc,-p_\cc,-p_\cc).
\EEQ
Electro-vacuum theory  also assumes a vacuum energy density $\rho_\vac(t,r)$  and its radial flux  $S_\vac(t,r)$, combined in
\BEQ
T_\vac^\mn=\rho_\vac \, g^\mn+S_\vac \,\cCc^\mn_{01}, \quad \cCc^\mn_{01}=\delta^\mu_0\delta^\nu_1+\delta^\mu_1\delta^\nu_0.
\EEQ
The full stress energy tensor contains the background and the cc's  material, electrostatic and vacuum terms, 
\BEQ \label{TtotMn=}
T^\mn =T_\cosm^\mn+T^\mn_\cc+T^\mn_{\em}+T^\mn_{\vac}  .
\EEQ

We express the Einstein equations as $\bar G^\mu_\ednu=\bar T^\mu_\ednu$ where
\BEQ \label{delTMn}
\hspace{-1mm}
\bar G^\mu_\ednu \equiv  \frac{G^\mu_\ednu}{8\pi G} -T_{\cosm \,\nu}^\mu  -T^\mu_{\cc\,\nu} ,
\quad
\bar T^\mu_\ednu=T^\mu_{\em \, \nu} \pplus T^\mu_{\vac \, \nu}  ,
\EEQ
with lowering of indices by $g_\mn$.
In the spirit of our earlier approaches, we assume that the FLRW background $T_{\cosm \,\nu}^\mu$ and the matter component  $T^\mu_{\cc\,\nu} $
are known, so we can express  $\bar\rho_E \equiv \bar G^\mu_{\ed \nu} C^\nu_{1\,\mu}/4$ and 
 $\bar\rho_\vac\equiv \bar G^\mu_{\ed \mu}/4$ in terms of $g_{0,1}$ and their derivatives.
The condition $\bar G^0_{\ed0}=\bar G^1_{\ed 1}$  leads to a quadratic equation for $g_1$ in terms of  $g_0$ and its derivatives. 
The results for  $\bar\rho_E$, $\bar\rho_\vac$ and $g_1$ are given in eqs. (\ref{barrhovac=})--(\ref{gw=}) of the Appendix.
Taking $S_\vac=G^1_{\ed 0}/8\pi G$ solves the last nontrivial components of the Einstein equations.

Given a  cosmological model for $\rho_0(t)$ and $p_0(t)$,  and the excess matter profiles $\rho_\cc(t,r)$ and $ p_\cc(t,r)$,
the remaining task is to solve $g_0(t,r)$ from $\bar\rho_E(t,r)=\rho_E(t,r)$, where
\BEQ \label{rhoEinex}
\rho_E^\inn
=\frac{q^2}{8\pi a^4g_0^2 R^2r^2} 
,\qquad 
\rho_E^\ext=\frac{q^2}{8\pi a^4 g_0^2 r^4}  ,
\EEQ
 in the interior and exterior, respectively.
Elimination of $g_1$ leads to a  lengthy fourth order partial differential equation for $g_0(t,r)$.
From the solution,  $\rho_\vac=\bar\rho_\vac$ and  $S_\vac$ can be read off,  showing that the vacuum is slaved by 
 electrostatics and matter, as happens in general.

The leading behavior at small $r$ is
\BEQ
\bar\rho_\vac = \bar\rho_E=\frac{1-g_1(t,0)}{16 \pi G a^2(t)g_0(t,0) }\frac{1}{r^2}+\cO(r^0) .
\EEQ
Consistent with  (\ref{rhoEinex}), it generally has a $1/r^2$ singularity.

  Eq. (\ref{v0=,q=}) yields  $q^2/R^2=v_0^2/G$ with $v_0\ll1$, hence we  set
\BEQ
g_{0,1}(t,r)=1-v_0^2 h_{0,1}(t,r),\quad \rho_\cc=v_0^2\rhovv ,\qad p_\cc=v_0^2 \pvv ,
\EEQ
and linearize in $v_0^2$. The resulting equations are given in eqs. (\ref{rhov-lin})--(\ref{h1-lin}) of the Appendix.

Since $\bar\rho_E$ contains the explicit factor $1/r^2$ of $\rho_E^\inn$ of (\ref{rhoEinex}), 
 $h_0 $  and $h_1$ are finite at $r=0$. We expand
\BEQ
 h_{0,1} (t,r) = h_{0,1}^{(0)}(t) + h_{0,1}^{(2)}(t) r^2 +h_{0,1}^{(4)}(t)r^4+\cdots
\EEQ
At $r=0$, the condition $\bar\rho_E/\rho_E^\inn=1$ yields $h_1^{(0)}=2/a^2$. Given a shape for $h_0^{(0)}(t)$, 
the expression (\ref{h1-lin}) for $h_1$  then yields $h_0^{(2)}(t)$. Likewise, order $r^2$ will set $h_0^{(4)}(t)$, and so on.
This provides a proper boundary condition for numerical integration from small $r$ up to $R$.

At large $r$, the Ansatz 
$h_{0}\approx k_4(t)/r^4+k_6(t)/r^6$, $h_{1}\approx l_4(t)/r^4+l_6(t)/r^6$  
solves $l_4$ in terms of $k_4$, $\dot k_4$ and $\ddot k_4$, 
so that $\bar\rho_E/\rho_E^\ext=1$ leads to 
a fourth order differential equation for $k_4(t)$ in terms of $a(t)$ and possibly
 $\sigma_4(t)$ and $\tau_4(t)$ from the asymptotics  $\rhovv \approx \sigma_4 /r^4+\sigma_6/r^6$,  
$\pvv \approx \tau_6 /r^4+\tau_6 /r^6$.  
Next order determines $k_6(t)$. 
One can thus solve $h_0$ from large $r$ down to $R$ and equate $h_0(t,R^+)$ to $h_0(t,R^-)$ by adjusting $h_0^{(0)}(t)$.
Hence the problem is uniquely posed, also beyond linearity. 
The resulting decay $S_\vac\sim 1/r^5$ confirms that no vacuum energy flows  in from infinity   into the 
  cocoon.

\subsection{Fit in the early Universe}
The above analysis affirms that the cocoons have  comoving size  $R\sim 2$ Mpc. 
The comoving  horizon passes $R$ for $z \sim 2.3 \, 10^5$ or $T\sim 55$ eV; 
at higher $z$ the cocoon is only partly in causal contact. 

With Planck redshift $z_P=\sqrt{1/G}\, / 2.725$ K = $5.2 \, 10^{31}$, the ccs have physical radius  $R_P=R/z_P\sim 1.2$ nm at the Big Bang.
The average charge per Planck volume $q(\ell_P/R_P)^3\sim 8.8 \, 10^{-23}e$ may relate to a unit of mini-charge.
We recall that in the Lorentz model for the electron, its charge stems from a shell of discrete or continuous mini-charges
\cite{lorentz1901scheinbare,poincare1905dynamique,nieuwenhuizen2024aether}. 

\section{Discussion}
Recently, Mistele et al.  \cite{mistele2024indefinitely} analyze the weak lensing of isolated galaxies.
The observations imply ``indefinite flattening of rotation curves'',   viz. the constancy of circular rotation speeds persists
 up to distances of 1 Mpc, the size of galaxy clusters. 
This  size of dark matter ``cocoons'' is much larger than expected from $\Lambda$CDM, and  difficult to explain within this theory.

The situation is modeled as ``charged cocoons'' within our recent electro-vacuum theory; the latter
 applies to vastly different scales from the Lorentz electron \cite{nieuwenhuizen2024aether} to
 black holes  \cite{nieuwenhuizen2023exact} and the whole  cosmos \cite{nieuwenhuizen2024solution}. 
In this theory,  dark matter does not stem from some unknown particle but is a combination of electrostatic and vacuum energy, with the
Coulomb repulsion of like charges stabilized by the Poincar\'e  stress (negative pressure) of the vacuum.
For the cocoons, a $\pm1/r^2$ charge density profile leads to a rotation profile that is basically flat up to 1 Mpc. 
The dark mass of the   cocoon coincides with the typical mass of a small galaxy cluster.
Different shapes of the charge profile can be considered.

On the Mpc scale,  the enclosed mass components occur at their cosmic fractions,  so that
the indirectly observed circular rotation speeds $v\sim 200$ km/s confirm the typical cocoon radius $R \sim  v /H_0\sim 2$ Mpc.

Like most galaxies, those with a charged cocoon likely contain a central supermassive black hole. 
Black holes created by stellar collapse inside charged cocoons are naturally charged.
Indeed, the cocoon's charge-to-mass ratio  $q/M\sqrt{G}\sim c/v_0 \sim 1500$  for $v_0$ $\sim 200$ km/s,
 is much larger than  the value 1 of an extremal black hole.
This supports physical realization of  our exact solutions for non-rotating, charged black holes with a regular core \cite{nieuwenhuizen2023exact}.

The cocoon size and mass, being comparable to that of galaxy clusters, hints at a common nature.
In this view, clusters involve several plus and minus charged cocoons. 
As mentioned  below (\ref{deltae=}), the charge mismatch factor $\delta_e$ in the solar neighborhood matches the estimates from ccs.
The smaller values in the fat clusters A1689 and A1835 may express a mixture of such cocoons of either charge.
In between the clusters,  there may be many charged cocoons.

The present-day analytical solution is generalized to the expanding Universe and expressed in a partial differential equation 
involving the FLRW background.

 That  the charged cocoons arise from merging of smaller ones is unlikely,
if only because these must  mainly have the  same sign of their net charges on the comoving Mpc scale.
If they neither arise from an instability, they must be present at the Big Bang, with  nanometer size in the Planck regime.
The involved inhomogeneous initial state
offers a new starting point for the  study of  the baryon asymmetry and  the $^7$Li problem.
In fact, the 1.2 nm physical cocoon size at the Big Bang  equals $\exp(60)$ Planck lengths $\ell_P=\sqrt{G}$, so
that after 60 $e$-folds  the inhomogeneous initial state  becomes homogeneous.
This may be related to the 60 $e$-folds expected for a graceful exit of inflation \cite{liddle2000cosmological}.

 The ``little red dots'', compact and extremely red galaxies  at  $z\ge 4$ \cite{akins2024cosmos}, may exhibit a similar connection to ccs. 

The existence of  Mpc sized cocoons predicts a peak in the cosmic microwave background at the arcmin scale or angular index $l\sim 8000$.
Both the South  Pole Telescope \cite{henning2018measurements}  (data up to $l=5000$)  and the Atacama Cosmology Telescope 
\cite{choi2020atacama}  (data up to $l=7500$) observe an upward trend in the temperature spectrum for $l\gtrsim 3000$.
While this has been explained in terms of cosmic foregrounds, it is interesting to test a connection to cocoons, 
and to extend the  observations towards the arcsec regime.

Finally one may speculate about the origin of  the induced nanometer  sized cocoons at the Big Bang. For instance,
 the collapse of the Universe in a previous aeon after all matter and remaining energy ended up in supermassive black holes,
 a case connecting to a cyclic Universe.

  



\onecolumn

\section{Appendix}

The expression for $\bar\rho_\vac(t,r)$ in terms of $a(t)$, $g_{0,1}(t,r) $, and $\rho_\cc(t,r)$ and $p_\cc(t,r)$  reads
\BEQ \label{barrhovac=}
 &&   \hspace{-8mm}
\bar\rho_\vac=\frac{1}{8\pi G}\Big [
\frac{1-g_0g_1}{2 a^2 g_0 r^2} 
-\frac{g_1'{}}{a^2 g_0r} -\frac{3 g_0'{} g_1}{2 a^2 g_0^2 r}
+\frac{\dot a^2}{4a^2}\left( \frac{6}{g_0 g_1}-2g_0 -\frac{g_0}{ g_1}- 3 g_0 g_1\right)
-\frac{3g_0'' g_1}{4a^2 g_0^2}+\frac{3 g_0'{}^2 g_1}{8 a^2 g_0^3}-\frac{3 g_0'{} g_1'{}}{4 a^2 g_0^2} -\frac{g_1''}{4 a^2g_0}
\qad \nn \\ &&   \hspace{-8mm}
+\frac{\ddot a}{2a}\left( \frac{3}{g_0 g_1}-2 g_0-\frac{g_0}{ g_1}\right)
+\frac{\dot a}{a}\frac{ 9 \dot g_0g_1-7  g_0 \dot g_1}{4  g_0^2 g_1^2}
+\frac{3\ddot g_0}{4 g_0^2 g_1}-\frac{\ddot g_1}{4 g_0  g_1^2}
-\frac{3 \dot g_0^2}{8 g_0^3 g_1}-\frac{3 \dot g_0 \dot g_1}{4 g_0^2 g_1^2}+\frac{\dot g_1^2}{2 g_0 g_1^3} \Big]
-\frac{ \rho_\cc-3p_\cc }{4}  .    
\EEQ
Likewise, the expression for $\bar\rho_E$ reads
\BEQ   \label{barrhoE=}
\hspace{-8mm}
\bar \rho_E &=& \frac{1}{8\pi G} \Big[
\frac{1-g_0g_1}{2 a^2 g_0 r^2} -\frac{g_0'{} g_1}{2 a^2 g_0^2 r}
+\frac{\dot a^2}{4a^2}\left(2g_0 -\frac{g_0}{ g_1}+ \frac{2}{g_0 g_1} -3 g_0 g_1\right)
  + \frac{1}{4a^2g_0}\big(\frac{g_0'' g_1}{g_0}-\frac{3g_0'{}^2 g_1}{2 g_0^2}+\frac{ g_0'{} g_1'{}}{g_0}
+ g_1'' \Big)
\nn \\ &&  \hspace{-5mm}
+\frac{\ddot a}{a}\left( g_0 - \frac{g_0}{ 2g_1}-\frac{1}{ 2g_0g_1}\right)
+\frac{3 \dot a  \dot g_1}{4 a g_0 g_1^2} +\frac{\dot a \dot g_0}{4a g_0^2g_1}
-\frac{\ddot g_0}{4 g_0^2 g_1} +\frac{\ddot g_1}{4 g_0 g_1^2} +\frac{3\dot g_0^2}{8 g_0^3 g_1}+\frac{\dot g_0 \dot g_1}{4 g_0^2 g_1^2}-\frac{\dot g_1^2}{2 g_0 g_1^3} \Big]
- \frac{\rho_\cc+p_\cc}{4}  .
\EEQ
The solution for $g_1$ reads
\BEQ \label{ge=}
g_1= \frac{\sqrt{ g_w  }-8 \pi  G (\rho_\cc+p_\cc) a^2  g_0^3 }{g_0 \left(6 \dot a ^2  g_0^4+2 g_0''  g_0-3g_0'{}^2\right)} ,
\EEQ
with
\BEQ \label{gw=}
g_w&=& 
2 \dot a ^2  \left(g_0^2+2\right) g_0^2 \left(6 \dot a ^2  g_0^4+2 g_0''  g_0-3g_0'{}^2\right)  
+a^2 \left(2 \ddot g_0 g_0-3 \dot g_0^2\right) \left[ 3g_0'{}^2-2 g_0 \left(3 \dot a ^2  g_0^3+g_0'' \right)\right]
\nn\\ 
&+& 2 a g_0 \left[2 \ddot a  g_0 \left(g_0^2-1\right)+\dot a  \dot g_0\right] \left(6 \dot a ^2  g_0^4+2 g_0''  g_0-3g_0'{}^2\right)
+ [8 \pi  G (\rho_\cc+p_\cc) a^2  g_0^3 \, ]^2 .
\EEQ
In the interior one has to solve $g_{0,1}(t,r)$ from  ${\bar \rho_E}={\rho_E^\inn}$ and in the exterior from ${\bar \rho_E} ={ \rho_E^\ext} $, where 
\BEQ \label{rhoEinex2}
\rho_E^\inn =\frac{v_0^2}{8\pi G a^4 g_0^2 r^2}  , \qquad \qquad \rho_E^\ext =\frac{v_0^2R^2}{8\pi G a^4 g_0^2 r^4}  . 
\EEQ

The expansion up to order $v_0^2$,
\BEQ
g_0(t,r)=1 - v_0^2h_0(t,r),\qquad g_1(t,r)=1 - v_0^2h_1(t,r), \qquad \rho_\cc(t,r)=v_0^2\rhovv (t,r),\qquad p_\cc(t,r) = v_0^2\pvv  (t,r), 
\EEQ
 yields the linearized  expressions 
\BEQ
 \label{rhov-lin}
\bar \rho_\vac =\frac{v_0^2}{8 \pi G }\Big[  
 \frac{ h_1}{2a^2r^2} + \frac{ 3h_0' +2h_1'}{2a^2r}  + \frac{3h_0''  \pplus h_1''}{4a^2}  
  - \frac{3\ddot h_0 \mmin  \ddot h_1}{4}
  + \frac{ \ddot a}{a}  (3h_0 \pplus h_1)  + \frac{ \dot a{} ^2}{a^2}    (3h_0 \pplus 2h_1)
   - \frac{\dot a{} }{a} \frac{9\dot h_0 \mmin 7 \dot h_1}{4}   \Big]  - v_0^2\frac{ \rhovv -3\pvv   }{4}   ,
\EEQ
and
\BEQ 
 \label{rhoE-lin}
\hspace{-6mm}
\bar \rho_E =\frac{v_0^2}{8 \pi G }\Big[  \frac{ h_1}{2a^2r^2} + \frac{ h_0'{}}{2a^2r} -  \frac{h_0'' +h_1'' }{4a^2}  + \frac{\ddot h_0-\ddot h_1}{4} 
- \frac{ \ddot a}{a}   (h_0+h_1) + \frac{ \dot a{} ^2}{a^2}    (h_0+h_1) - \frac{\dot a{} }{a} \frac{\dot h_0+ 3 \dot h_1}{4}  \Big] 
-v_0^2 \frac{\rhovv +\pvv  }{4}   .
\EEQ
The linearized version of (\ref{ge=}), (\ref{gw=})  reads 
\BEQ \label{h1-lin}
h_1=
-\frac{a^2}{6 \dot a{} ^2} \ddot h_0 +\frac{2a\ddot a }{3 \dot a{} ^2}  h_0 +\frac{a }{6 \dot a{} } \dot h_0 
-\frac{2}{3}h_0 -\frac{h_0''}{6 \dot a{} ^2} + \frac{a^2}{6 \dot a{} ^2} (\rhovv +\pvv  )  .
\EEQ

To order $v_0^2$, eq. (\ref{rhoEinex2})  reduces to
\BEQ
\rho_E^\inn =\frac{v_0^2}{8\pi G a^4 r^2}  , \qquad \qquad \rho_E^\ext =\frac{v_0^2R^2}{8\pi G a^4 r^4}  . 
\EEQ

\end{document}